\documentstyle[11pt]{article}
 \let\msk=\medskip \let\bsk=\bigskip
\let\qd=\quad  

\let\a=\alpha \let\be=\beta  \let\de=\delta
\let\ep=\varepsilon  \let\h=\eta
  \let\la=\lambda \let\m=\mu
\let\n=\nu  \let\r=\rho 
\let\om=\omega 
\let\ph=\varphi  \let\Ps=\Psi
\let\Om=\Omega  
 \let\Ga=\Gamma \let\De=\Delta

\def\0#1#2{\frac{#1}{#2}} \def\s0#1#2{\mbox{\small{$\frac{#1}{#2}$}}}
\def\2{{\times}} \def\3{\vec }
\def\5{\bar }  \def\6{\partial } \def\7{\hat } \def\4{\tilde }
  
 \let\LRA=\Leftrightarrow
\let\then=\Rightarrow 

\let\nn=\nonumber
\def\bea{\begin{eqnarray}} \def\eea{\end{eqnarray}}
\def\beann{\begin{eqnarray*}} \def\eeann{\end{eqnarray*}}
\def\beq{\begin{equation}} \def\eeq{\end{equation}}
\def\ba{\begin{array}} \def\ea{\end{array}}
\def\ben{\begin{enumerate}} \def\een{\end{enumerate}}

 \def\cb{{\cal B}} \def\cg{{\cal G}}
\def\ca{{\cal A}} \def\cd{{\cal D}} \def\cl{{\cal L}}
 \def\cN{{\cal N}} 
\def\cF{{\cal F}}  
\def\cW{{\cal W}}  
\def\cC{{\cal C}}

 \def\R#1#2#3{{R_{#1#2}}^{#3}}
\def\f#1#2#3{{f_{#1#2}}^{#3}}
\def\cf#1#2#3{{{\cal F}_{#1#2}}^{#3}}
\def\cA#1#2{{{\cal A}_{#1}}^{#2}}
 \def\o#1#2{\om_{#1}{}^{#2}}
\def\e#1#2{e_{#1}{}^{#2}} \def\E#1#2{E_{#1}{}^{#2}}
\def\bg#1#2{\7e_{#1}{}^{#2}} \def\et#1#2{h_{#1}{}^{#2}}

\def\Npar{N_\6} \def\on{{\5n}}

\def\csum#1#2{\sum_{#1}\hspace{-1.#2em}\circ\ \ \ }

\newcommand{\mysection}[1]{\section{#1}
            \setcounter{equation}{0}\setcounter{figure}{0}}
\def\Gl#1{(\ref{#1})} \def\gh{\mbox{gh}\, }
\def\PH{{\cal T}}
\def\Th{\4C}
\def\mod{\mbox{mod}}
\begin{document}
\hspace*{\fill} NIKHEF-H 93--21\\
\begin{center}{\LARGE {\bf Structure of BRS-Invariant Local Functionals
}}\vspace{.5cm}

{\renewcommand{\thefootnote}{\fnsymbol{footnote}}
{\Large Friedemann Brandt}\footnote{Supported by Deutsche
Forschungsgemeinschaft}}
\setcounter{footnote}{0}\vspace{.5cm}

NIKHEF-H, Postbus 41882,
1009 DB Amsterdam, The Netherlands\end{center}

\begin{abstract}
For a large class of gauge theories a nilpotent BRS-operator $s$ is constructed
and its cohomology in the space of local functionals of the off-shell
fields is shown to be isomorphic to the cohomology of
$\4s=s+d$ on functions $f(\4C,\PH)$ of tensor fields $\PH$
and of variables $\4C$ which are constructed of the
ghosts and the connection forms. The result allows general statements about
the structure of invariant classical actions and anomaly candidates
whose BRS-variation vanishes off-shell.
The assumptions under which the result holds
are thoroughly discussed.
\end{abstract}

\mysection{Introduction}\label{intro}

The BRS-formalism \cite{brs,bau} is
an elegant and powerful tool for dealing with gauge symmetries
in quantum field theory. It is
available for a large class of gauge theories
whose classical symmetries have an algebra which closes on the off-shell
fields. In particular the BRS-formalism allows
to characterize classical actions and anomalies as
BRS-invariant local functionals of the fields. Namely an
action is a BRS-invariant functional with ghost number 0 and
anomalies correspond to BRS-invariant functionals with ghost number 1.
The latter can occur in
the quantized theory if there is no regularization preserving all
symmetries of the classical theory. The BRS-invariance of the classical
action follows from its gauge invariance (and vice versa)
and BRS-trivial contributions to the action
can be used to construct elegantly a gauge fixing and the corresponding
Faddeev-Popov ghost contributions to the action \cite{bau}. The
BRS-invariance of anomalies comprises consistency conditions
which anomalies have to satisfy as a consequence of the algebra of the
symmetries of the classical theory and which have been first derived for
Yang--Mills theories \cite{wz}. In summary, actions and anomalies are
solutions of the so-called consistency equation
\beq s\, \cW^G=0,\qd \gh (\cW^G)=G\label{1}\eeq
where $s$ denotes the BRS-operator and $\cW^G$ is a local functional with
ghost number ($\gh$) $G$. In this paper I restrict the
investigation to the case that \Gl1 holds identically in the fields
(i.e. for off-shell fields). I remark however that according to
recent results \cite{proe,anti} there might be also anomalies
corresponding to on-shell solutions of \Gl1. More precisely these functionals
satisfy \Gl1 only weakly in the sense that they are BRS-invariant
only up to functionals which vanish for solutions of
the classical field equations.

The above-mentioned locality of $\cW^G$ is in the case
of an invariant action $\cW^0$ a physical
requirement (input) and follows in the case of an anomaly $\cW^1$ from
renormalization theory. Here a functional is called local if its
integrand is a formal (not necessarily finite) power series in the
undifferentiated elementary fields and a polynomial in their partial
derivatives. In addition the integrand may depend explicitly on the coordinates
(e.g. via background fields). The nilpotency of the BRS-operator (see below)
implies that \Gl1 represents a cohomological problem, i.e. in order to
solve \Gl1 one has to determine the cohomology of $s$ in the space of local
functionals with ghost number $G$. In particular each functional of the form
$sX^{G-1}$ solves \Gl1 and is therefore called a trivial solution
provided $X^{G-1}$ is a local functional. Two solutions of
\Gl1 are called equivalent if they differ by a trivial solution:
\beq \cW^G\cong\cW'{}^G\qd\LRA\qd\cW^G-\cW'{}^G=s\,X^{G-1}.\label{3}\eeq
A local functional of the fields is an
integrated local volume form $\om_D^G$ of the fields
(the superscript of a form denotes its ghost number, the subscript its
form--degree, $D$ denotes the space-time dimension)
\[ \cW^G=\int\om_D^G \]
and BRS--invariance of $\cW^G$ requires its integrand to satisfy
\beq s\,\om_D^G+d\,\om^{G+1}_{D-1}=0\label{i1}\eeq
where $\om_{D-1}^{G+1}$ is a local
$(D-1)$--form with ghost number $G+1$ and $d$ denotes the exterior derivative
\beq d=dx^m\6_m.\label{ext}\eeq
The BRS-operator is by construction a
nilpotent antiderivation which anticommutes with $d$:
\beq s^2=\{s,d\}=d^2=0.\label{i2}\eeq
Therefore each volume form which can be written as
$s\,\h_D^{G-1}+d\,\h_{D-1}^G$ solves (\ref{i1}). Such
solutions of (\ref{i1}) are called trivial provided $\h_D^{G-1}$ and
$\h_{D-1}^G$ depend locally on the fields. Analogously to \Gl3
two solutions of (\ref{i1}) are called equivalent if they differ
by a trivial solution,
\beq \om_D^G\cong\om'{}^G_D\qd\LRA\qd
\om'{}^G_D-\om_D^G=s\,\h_{D}^{G-1}+d\,\h_{D-1}^G. \label{i3}\eeq
\Gl1 resp. \Gl{i1} have been investigated by many authors for various theories.
In particular \Gl{i1} has been solved completely
for Yang--Mills theories \cite{com} where the solutions
can be divided into two classes: `Lagrangian solutions' constructed of
tensor fields and
undifferentiated ghosts and `chiral solutions' which can be written
completely in terms of 0--forms
given by the ghosts, connection 1--forms constructed of the gauge fields and
the corresponding curvature 2--forms. Examples for
`Lagrangian solutions' are invariant Lagrangians (times the volume
element $d^Dx$), examples of `chiral solutions' are Chern-Simons forms and the
integrands of chiral anomalies (see e.g. \cite{tal,zu}).
Analogous results hold for Einstein--Yang--Mills
theories in an Poincar\'e-invariant background \cite{grav}.
The present paper shows that the results for
Yang--Mills and Einstein--Yang--Mills theories
hold in a generalized form for a large class of gauge theories.
We shall see however that
in the general case it does not make much sense to distinguish
`Lagrangian' and `chiral solutions'---both arise from the same
universal structure of solutions. The derivation of this structure
is the main result of this paper.
It will become clear that it originates in the very definition
of the gauge theories investigated here and is intimately related with
a generalized tensor calculus. Therefore this definition
will be thoroughly discussed. It will be given in an approach which treats
all gauge symmetries on an equal footing and in particular
takes into consideration diffeomorphisms
`automatically'. As a consequence the latter
always contribute to the BRS-transformations. Therefore ordinary
Yang--Mills theories are strictly speaking not covered by the investigation
of this paper, unlike Einstein--Yang--Mills theories. Other prominent
theories to which the results of this paper apply are
supergravity theories \cite{sugra}.

The paper is organized as follows. Section \ref{ass} defines the class
of gauge theories for which the results of this paper hold, including
the construction of the BRS--operator. Section \ref{res} contains the result
of the investigation, section \ref{proof} sketches its proof.
Sect. \ref{rem} comments on the importance of some underlying assumptions
for the validity of the result and in particular discusses modifications
of the result if these assumptions are not fulfilled.
The paper is completed by a summary and an appendix which contains
the proof of two lemmas used in section \ref{proof}.

\mysection{Assumptions}\label{ass}

\subsection{Algebra on tensor fields}\label{ten}

The gauge theories treated in this paper can be characterized
starting from an algebra of operators $\De_M$ which is realized
on tensor fields. I postpone a precise definition of tensor fields
to subsection \ref{var} but I remark that a tensor field is a
function $\PH(x,\ph,\6\ph,\6\6\ph,\ldots)$ depending on
the elementary fields and their partial derivatives (and explicitly on the
coordinates via the background vielbein, see below) such that the
BRS--transformation of $\PH$ has a special simple form. The operators
$\De_M$ whose number is not necessarily finite
 are assumed to be independent and to have a closed algebra
on tensor fields which is of the form
\beq [\De_M,\De_N\}:=\De_M\De_N-(-)^{|M|\, |N|}\De_N\De_M=\cf MNP\De_P
,\qd |M|,|N|\in\{0,1\}
\label{a6}\eeq
where $|M|$ denotes the grading of $\De_M$
(i.e. we allow for bosonic and fermionic $\De$'s) and the
structure functions $\cf MNP$ are generally field dependent,
i.e. they are tensor fields themselves.
According to \Gl{a6} they
are symmetric or antisymmetric in their lower indices and even or odd
graded depending on the grading
of the corresponding generators
\beq \cf MNP=-(-)^{|M|\, |N|}\cf NMP,\qd |\cf MNP|=|M|+|N|+|P|
\qd \mod\ 2.
\label{t2}\eeq
By assumption they satisfy the Jacobi- resp. Bianchi-identities
\beq \csum {MNP}{5} (-)^{|M|\, |P|}\left(\De_M\cf NPQ-\cf MNR\cf RPQ\right)=0
\label{a6a}\eeq
which guarantee the consistency of the algebra \Gl{a6}. The
sum in (\ref{a6a}) is the cyclic sum, i.e.
\beq\csum {MNP}{5}(-)^{|M|\, |P|}X_{MNP}=(-)^{|M||P|}X_{MNP}+
(-)^{|N||M|}X_{NPM}+(-)^{|P||N|}X_{PMN}.
\label{t4}\eeq
Notice that by assumption the operators $\De_M$ are covariant in the
sense that they map tensor fields to tensor fields.
It is not assumed that they
can be defined on all fields in a way that
\Gl{a6} holds. In particular they are generally not defined on
the gauge fields $\cA mN$ and the ghost fields $\7C^N$ which can be
introduced elegantly by requiring that
the exterior derivative $d$ and the BRS--operator $s$ act on
tensor fields $\PH$ according to
\bea  d\, \PH&=&\ca^N\De_N\PH,\label{ten9a}\\
 s\, \PH&=&\7C^N\De_N\PH\label{ten9}\eea
where $\ca^N$ denotes the gauge field 1--form
\beq  \ca^N=dx^m\cA mN.\label{ten9b}\eeq
Inserting $d=dx^m\6_m$ and (\ref{ten9b}) into \Gl{ten9a} one obtains
\beq \6_m\PH=\cA mM\De_M\PH.\label{ten9c}\eeq
The grading of the ghosts, gauge fields and differentials is fixed
by the requirement that $s$ and $d$
are odd graded operators while $\6_m$ is even graded. This gives
\beq |\7C^N|=|N|+1\qd \mod\ 2,\qd |\cA mN|=|N|,\qd |dx^m|=1.\label{ten9d}\eeq
According to \Gl{ten9a} and \Gl{ten9}
both $d$ and $s$ can be written on tensor fields
as linear combinations of the operators
$\De_N$ where the coefficients are the gauge field 1--forms $\ca^N$
in one case and the ghosts $\7C^N$ in the other case. This
similarity of $s$ and $d$ is one of the decisive reasons behind the
result of this paper and therefore it is commented now.
\Gl{ten9} of course is the definition of
the BRS--operator on tensor fields and at the same time serves
as the defining property of tensor fields
(see subsection \ref{var}). \Gl{ten9a} can be viewed as the definition
of the covariant derivatives. Namely the latter form
by assumption a subset of $\{\De_M\}$. In order to distinguish the
covariant derivatives from the remaining $\De$'s we label the former
by latin indices from the beginning of the alphabet
and the latter by greek indices
from the middle of the alphabet. Furthermore the more
customary notation $\cd_a$ is introduced for the covariant derivatives:
\[ \{\De_M\}=\{\cd_a,\De_\m\},\qd \cd_a\equiv\De_a,\qd a=1,\ldots,D,\qd
\m=D+1,D+2,\ldots \]
The corresponding gauge and ghost fields are denoted by $\cA ma$,
$\cA m\m$, $\7C^a$ and $\7C^\m$. The covariant derivatives of course are
by assumption bosonic (even graded) operators:
\beq |a|=0.\label{ten9e}\eeq
The special role played by the covariant derivatives originates in the
important assumption that
the gauge fields $\cA ma$ form (at each point) an invertible
$D\times D$--matrix which is called the vielbein.
We shall therefore also use the more customary notation
$\e ma$ instead of $\cA ma$. The entries
of the inverse vielbein are denoted by $\E am$:
\beq \cA ma\equiv\e ma,\qd
\e ma\E an=\de_m^n,\qd \E am\e mb=\de_a^b.\label{a3}\eeq
The invertibility of the vielbein allows to solve (\ref{ten9c}) for $\cd_a\PH$
and thus \Gl{ten9a} indeed can be regarded as the definition of the
covariant derivatives according to
\beq \cd_a\, \PH=\E am(\6_m-\cA m\m\De_\m)\, \PH\label{a8}\eeq
which is familiar
from the Einstein--Yang--Mills theory where the $\De_\m$ are the elements
of a Liealgebra. Notice however that this restrictive assumption has been
dropped and replaced by the more general requirement that the $\De_\m$
constitute together with the $\cd_a$ a set of
independent (generally graded) operators which have a closed algebra
\Gl{a6} on tensor fields.
In addition the $\De_M$ are assumed to be local
operators in the sense that each of them maps local tensor fields to
local tensor fields.

\subsection{BRS--operator and field strengths}\label{brs}

So far three kinds of fields have been introduced, namely tensor, gauge
and ghost fields. In order to be able to
construct a gauge fixed BRS-invariant action one introduces
in addition an antighost $\zeta^N$ and a `Lagrange multiplier field' $b^N$
for each gauge field with gradings given by
\[|\zeta^N|=|N|+1\qd \mod\ 2,\qd |b^N|=|N|.\]
On tensor fields the BRS--operator has been defined already in \Gl{ten9}.
In addition we define the BRS--transformations of the antighosts, explicit
coordinates and differentials according to\footnote{The fields
and their partial derivatives
are not regarded as functions of variables $x$ but as fundamental
variables themselves, see subsection \ref{var}.}
\beq s\, \zeta^N=b^N,\qd s\, x^m=0,\qd s\, dx^m=0\label{b2}\eeq
and further require that
\beq (s+d)^2=0\qd\LRA\qd s^2=[s,\6_m]=[\6_m,\6_n]=0.\label{b1a}\eeq
This implies in particular
\beq s\, \6_{m_1}\ldots\6_{m_k}\zeta^N=\6_{m_1}\ldots\6_{m_k}b^N,\qd
s\, \6_{m_1}\ldots\6_{m_k}b^N=0.\label{b2a} \eeq
The BRS--transformations of the ghosts and gauge
fields follow from
requiring (\ref{b1a}) to hold on tensor fields. Notice that
\Gl{ten9a} and \Gl{ten9} imply
\beq (s+d)\, \PH=\4C^N \De_N\, \PH \label{b4}\eeq
where
\beq \4C^N=\7C^N+\ca^N.\label{b5}\eeq
$(s+d)^2\PH$ can be evaluated using \Gl{a6} and requiring it to vanish
one obtains
\beq (s+d)\,\4C^P=\s0 12(-)^{|N|} \4C^N \4C^M\cf MNP\label{b6}\eeq
which reads more explicitly
\bea
s\,  \7C^P  &=&\s0 12(-)^{|N|} \7C^N \7C^M\cf MNP,
                                                        \label{t17}\\
s\, \cA mP&=&\6_m \7C^P+ \7C^M\cA mN\cf NMP,
                                                        \label{t18}\\
0         &=&\6_m\cA nP-\6_n\cA mP-\cA mM\cA nN\cf NMP.
                                                        \label{t19}\eea
Notice that (\ref{b6}) does not only
define the BRS--transformation of the ghosts and
gauge fields \Gl{t17}, \Gl{t18} but also
contains the identity \Gl{t19} which guarantees the consistency of
(\ref{ten9c}) and $[\6_m,\6_n]\PH=0$. \Gl{t19}
 is not a differential equation restricting the
gauge fields and the structure functions but determines the structure
functions $\cf abN$ in terms of the gauge fields, their partial derivatives
and the remaining structure functions $\cf \m{\n}N$ and
$\cf \m{a}N$. Namely \Gl{t19}
can be solved for $\cf abN$ due to the invertibility of the vielbein:
\beq \cf abN=-\E am\E bn(\6_m\cA nN-\6_n\cA mN+\e nc\cA m\m\cf \m{c}N
-\e mc\cA n\m\cf \m{c}N+\cA n\n\cA m\m\cf \m{\n}N).\label{a13}\eeq
(\ref{a13}) justifies to call the $\cf abN$ the field strengths corresponding
to $\cA mN$. Notice that they
generally do not only depend on the gauge fields and their derivatives
but also on additional fields
which contribute via the structure functions $\cf \m{\n}N$ and
$\cf \m{a}N$ to the
r.h.s. of (\ref{a13}). In Einstein--Yang--Mills theory all these
structure functions are constant and the field strengths therefore depend
only on the gauge fields but, for instance, in supergravity theories this
is not the case and (\ref{a13}) yields in this case
the correct extension of the Yang--Mills field strengths
and the Riemann tensor and defines the field strengths of the
gravitino such that these fields transform covariantly
under general coordinate, Yang--Mills, Lorentz and
supersymmetry transformations.

Using (\ref{b4}) and (\ref{b6}) one can check that
(\ref{b1a}) holds also for the ghosts and the gauge fields
by virtue of the Jacobi- and Bianchi-identities \Gl{a6a}.

It will turn out that
the solutions of (\ref{i1}) can be more easily written in terms of
new ghost variables $C^m$, $C^\m$ than in terms of the ghosts $\7C^M$.
The different ghost basis' are related by
\beq \7C^a=\e ma C^m,\qd \7C^\m=C^\m+C^m\cA m\m.             \label{g11}\eeq
The new ghost basis is chosen such that the BRS--transformation of a tensor
field now takes the form
\beq s\, \PH=(C^m\6_m+C^\m\De_\m)\, \PH\label{ten20a}\eeq
as can be easily verified by inserting (\ref{a8}) into \Gl{ten9}.
In terms of the new ghost variables the BRS--transformations of the ghosts
and gauge fields read
\bea
s\, \e ma &=&C^n\6_n\e ma+(\6_mC^n)\, \e na+C^\m\cA mN\cf N\m{a},
                                                            \label{t23}\\
s\,\cA m\m&=&C^n\6_n\cA m\m+(\6_mC^n)\, \cA n\m+\6_mC^\m
+C^\n\cA mN\cf N\n\m,
                                                            \label{t24}\\
s\, C^m   &=&C^n\6_nC^m+\s0 12(-)^{|\m|}C^\m C^\n\cf \n\m{a}\E am,
                                                            \label{t25}\\
s\, C^\m  &=&C^n\6_nC^\m+\s0 12(-)^{|\n|}C^\n C^\r
             (\cf \r\n\m-\cf \r\n{a}\E am\cA m\m).
                                                            \label{t26}\eea
Of course this version of the BRS--algebra is
completely equivalent to the version using the $\7C^M$ due to the invertibility
of the vielbein.

\subsection{Field content, variables and tensor fields}\label{var}

We restrict the investigation to theories whose field content can be
characterized by means of the fields introduced so far. Thus the
`classical' field content consists of the components $\e ma$ of the vielbein,
the set of gauge fields $\cA m\m$ and a set of further fields
$\Ps^i$ which by assumption are tensor fields. The $\Ps^i$ are therefore
called elementary tensor fields. The field content is completed
by the ghosts $C^N$, the antighosts $\zeta^N$ and the Lagrange multiplier
fields $b^N$.
By this assumption we exclude for instance theories which contain
gauge potentials $\cb=dx^{m_1}\ldots dx^{m_p}\cb_{m_1\ldots m_p}$ but
it is expected that the presence of such fields (and corresponding
ghosts and ghosts of ghosts) leads only to small modifications
of the result, at least in cases of interest
(cf. the discussion of new minimal supergravity given in \cite{sugra}).

However some additional remarks are in order.
First it is stressed that
not the fields $\e ma$ themselves are regarded
as elementary fields but their deviations $\et ma$ from the entries $\bg ma(x)$
of some background vielbein which is also assumed to be invertible (but else
arbitrary)
\beq \e ma=\bg ma(x)+\et ma.\label{a2}\eeq
We allow that some (or all) gauge fields $\cA m\m$ are not
elementary fields but local functions of the fields $\et ma$, $\Ps^i$
and the remaining gauge fields. Gauge fields $\cA m\m$ which
are not elementary fields are called {\it composite} gauge fields in the
following\footnote{An example for a composite
gauge field is the spin--connection in gravity or supergravity
theories with vanishing structure function (torsion) $\cf abc$.}.
I stress however that compositeness
of gauge fields is required to be consistent with eqs. \Gl{t17}--\Gl{t19}
resp. \Gl{t23}--\Gl{t26} which
must hold identically in the elementary fields and their partial derivatives.
In particular these equations must not
impose differential equations for the ghosts.
Namely it is assumed that $\et ma$, $\Ps^i$, $C^m$ ,$C^\m$, $\zeta^N$, $b^N$
and a subset of $\{\cA m\m\}$
form a complete set of {\it elementary fields} $\{\ph^\a\}$
which means that there are no algebraic identities relating
the variables
\beq \6_{m_1}\ldots\6_{m_k}\ph^\a,\qd m_{i+1}\geq m_i, \qd
k\geq 0\label{a1}\eeq
apart from those which follow from their grading according to \Gl{grad}.
Thus we regard \Gl{a1}
as a set of infinitely many independent variables on which
the `partial derivatives' $\6_m$ act algebraically
(for instance $\6_m$ maps the variable $\ph^\a$ to the
variable $\6_m\ph^\a$, and $\6_m\6_n\ph^\a$ and $\6_n\6_m\ph^\a$ are
regarded as the same variable)\footnote{This approach can be formalized
using the jet bundle theory, see e.g. \cite{olv}.}.
Notice that here it is essential
that all fields are off-shell fields. Explicit coordinates
$x^m$ and the differentials $dx^m$ are treated as additional independent
variables on which the $\6_m$ act according to
\[ \6_m\, x^n=\de_m^n,\qd \6_m\, dx^n=0. \]
Notice that these
assumptions exclude cases where constant ghosts (corresponding e.g.
to global symmetries) are contained in $s$ though these cases may be
treated on a completely equal footing with the case where constant ghosts are
absent. Nevertheless the presence of constant ghosts
complicates the investigation as we shall see in section \ref{rem}.

Each variable has a definite grading which determines its statistics
according to
\bea & & z^Az^B=(-)^{|z^A|\,|z^B|}z^Bz^A,\qd
 z^A\in\{\6_{m_1}\ldots\6_{m_k}\ph^\a,\ x^m,\ dx^m\},\nn\\
& &|\6_{m_1}\ldots\6_{m_k}\ph^\a|=|\ph^\a|,\qd |x^m|=0,\qd |dx^m|=1.
\label{grad}\eea

I conclude this section with some remarks on tensor fields.
In subsection \ref{ten} tensor fields have been characterized by the property
that the operators $\De_M$ are realized on them according to
the algebra \Gl{a6}. In particular this allowed to
define the BRS--transformation of tensor fields according to \Gl{ten9}.
One may ask whether it is possible to characterize tensor fields more
concretely. The answer is yes and in fact surprisingly simple if we
now use \Gl{ten9} as a defining property for a function of the fields
to be a tensor field. More precisely we require that a tensor field
is a local function $\PH$ of $x^m$, $\et ma$, $\Ps^i$, $\cA m\m$ such that
$s\PH$ does not contain partial derivatives of the ghosts.
Using this definition one can {\it prove} that each tensor field
depends on the fields and their partial derivatives only via
the elements $\PH^r$ of
\beq\{ \PH^r\}=\{\cd_{a_1}\ldots\cd_{a_k}\Ps^i,\
\cd_{a_1}\ldots\cd_{a_k}\cf abN:\ k\geq 0\}\label{a5}\eeq
where $\cf abN$ is given by (\ref{a13}).
In order to prove this statement one uses the fact that according to
appendix \ref{app1}
each function of $x^m$, $\et ma$, $\cA m\m$, $\Ps^i$ and their partial
derivatives can be written also in terms of the variables
$x^m$, $\PH^r$ and the elements of
\[ \{U_l\}=\{\et ma,\ \cA m\m,\ \6_{(m_k}\ldots\6_{m_1}\e {m_0)}a,\
\6_{(m_k}\ldots\6_{m_1}\cA {m_0)}\m:\ k\geq 1\}\]
which is a subset of $\{u_l\}$ defined in appendix \ref{app1}.
The requirement that the BRS--transformation of a local function
$f(x,\PH,U)$ does not depend on derivatives of the $\7C^M$ is then
easily seen to require that $f$ does not depend on the $U_l$ at
all\footnote{I stress that this holds due to the assumption that
the ghosts are elementary fields.}.

Thus each tensor field can be written as a function of the $\PH^r$
which of course are tensor fields themselves, i.e. we know all tensor fields
constructable of the fields and their partial derivatives. Notice that
the $\PH^r$ are `created' by acting with $\cd_a$ on $\Ps^i$ and
$\cf abN$. Therefore the $\cd_a$ are in a sense
not only a subset of $\{\De_M\}$ but
also generate the space on which the $\De_M$ are realized, similarly
as the $\6_m$ generate the variables (\ref{a1}).
Notice that the invertibility of the vielbein is
responsible for this special part played by the $\cd_a$
compared to the remaining $\De$'s since it relates
the set $\{\cd_a,\De_\m\}$ of operators to the set $\{\6_m,\De_\m\}$
(on tensor fields).

Finally I remark that the definition of tensor fields given above requires
that each tensor field transforms scalarly
under general coordinate transformations since otherwise its
BRS--transformation would contain partial derivatives of the
ghosts $C^m$ of diffeomorphisms. This definition represents no
loss of generality since `world indices' $m,n,\ldots$ which
indicate a nonscalar behaviour under general coordinate
transformations can always be converted
to `flat indices' $a,b,\ldots$ by means of the vielbein and its
inverse (here of course the invertibility of the vielbein again
is essential).
Notice however that this in particular means that neither the vielbein (or
a metric $g_{mn}$ built from it) is considered as a
tensor field nor, for instance, the quantities $\cf mnN:=\e ma\e nb\cf abN$.

\mysection{Result}\label{res}

In order to formulate the result of this paper I define the cohomology of
\beq\4s=s+d\label{tildes}\eeq
 on local functions
\beq f^g(\4C,\PH)=\4C^{N_1}\ldots\4C^{N_g}f_{N_1\ldots N_g}(\PH).
\label{r-1}\eeq
This cohomology is denoted by $H^g(\4s)$. It is well-defined since $\4s$ leaves
the space of functions $f(\4C,\PH)$ invariant, i.e. the $\4s$-variation
of a function $f(\4C,\PH)$ can be always written completely in terms of the
variables $\4C^N$, $\PH^r$ due to \Gl{b4} and \Gl{b6}.
Recalling basic properties of cohomological problems I remark that
the general solution of
\beq \4s\, f^g(\4C,\PH)=0\label{r1}\eeq
is a linear combination of solutions $f^g_i$ which, loosely speaking, form
a basis for the solutions of \Gl{r1}.
More precisely $\{f^g_i\}$ is a set of solutions of \Gl{r1} which represent the
cohomology classes of $H^g(\4s)$:
\beann & &f^g(\4C,\PH)\cong
\sum_i a_i\, f^g_i(\4C,\PH),\qd \4s\, f^g_i(\4C,\PH)=0,\\
& &\sum_i a_i\, f^g_i(\4C,\PH)\cong 0\qd \LRA\qd a_i=0\qd\forall\, i
\eeann
where $a_i$ are constant coefficients ($c$-numbers) and
two solutions of \Gl{r1} are called equivalent if they differ only
by a trivial solution or a constant piece:
\beq f^g(\4C,\PH)\cong f'{}^g(\4C,\PH)\qd\LRA\qd f'{}^g(\4C,\PH)-f^g(\4C,\PH)
=\4s\, h^{g-1}(\4C,\PH)+const.\, .\label{r2}\eeq
I remark that the inclusion of the constant in (\ref{r2}) in our case serves
to treat elegantly the fact that $c$-numbers are solutions of (\ref{r1})
with $g=0$ (and in fact they are the only solutions with $g=0$ according
to the corollary given below). Notice that a constant can
contribute to $f^g(\4C,\PH)$ only for $g=0$ as a consequence of our
assumption that constant ghosts are absent. On the other hand there
are no contributions $\4s\, h^{-1}$ in the case $g=0$ since there are no
functions \Gl{r-1} for $g<0$. Thus $f^g\cong f'^g$ in fact holds in the case
$g=0$ iff $f^0- f'^0=const.$ and in the case $g>0$ iff $f'{}^g-f^g
=\4s\, h^{g-1}$. I stress however that in presence of constant
ghosts the inclusion of constants is less trivial.

Furthermore we define $H^{n,p}(s|d)$ as the cohomology of $s$ modulo
$d$ on local $p$--forms with ghost number $n$. This cohomology is
defined by the problem
\beq s\om_p^n+d\om_{p-1}^{n+1}=0,\qd \om_p^n\cong\om'{}_p^n\ \LRA\
\om_p^n-\om'{}_p^n=s\h_p^{n-1}+d\h_{p-1}^n\label{rr}\eeq
where $\om_p^n$, $\om_{p-1}^{n+1}$, $\h_p^{n-1}$ and $\h_{p-1}^n$
are local forms whose subscript (superscript) denotes their form degree
(ghost number). Again the general solution of \Gl{rr} can be written as
\beann \om_p^n\cong \sum_i a_i\, (\om_p^n)_i\eeann
where $\{(\om_p^n)_i\}$ denotes a set of solutions of \Gl{rr} which
represent the cohomology classes of $H^{n,p}(s|d)$. In particular
in order to solve \Gl{i1} one has to determine $H^{G,D}(s|d)$, i.e.
one has to find a set $\{(\om_D^G)_i\}$. The main result of the present
paper is the following:\msk

{\bf Theorem:} {\it In contractible manifolds the cohomologies
$H^{G,D}(s|d)$ and $H^{G+D}(\4s)$ are isomorphic, i.e. the cohomology
classes of $H^{G,D}(s|d)$ correspond one-to-one to those of $H^{G+D}(\4s)$
provided there are no constant ghosts. This correspondence is very explicit:
If $f_i^{G+D}(\4C^a,\4C^\m,\PH)$ is a solution of (\ref{r1})
representing a cohomology class of $H^{G+D}(\4s)$ then the
corresponding cohomology class of $H^{G,D}(s|d)$ is represented by a
nontrivial solution $(\om_D^G)_i$ of (\ref{i1}) given by
\beq (\om_D^G)_i=\left[ f^{G+D}_i(e^a,C^\m+\ca^\m,\PH)\right]_D,\qd
e^a=dx^m\e ma,\qd \ca^\m=dx^m\cA m\m\label{r0}\eeq
where $[ f^{G+D}_i(e^a,C^\m+\ca^\m,\PH)]_D$ denotes
the volume form contained in $f^{G+D}_i(e^a,C^\m+\ca^\m,\PH)$.
Thus the solutions of (\ref{i1}) can be easily obtained
from those of (\ref{r1}).}\msk

This theorem will now be spelled out in some more detail in order to
comment it and give some insight into its origin. Each function
$f^{G+D}(\4C,\PH)$ decomposes according to
\beq f^{G+D}(\4C,\PH)=\sum_{p=0}^D\om_p^{G+D-p}\label{dec}\eeq
where $\om_p^{G+D-p}$ is a $p$--form with ghost number $(G+D-p)$.
\Gl{r1} decomposes into the so-called descent equations
\beq 0<p\leq D:\qd s\om_p^{G+D-p}+d\om_{p-1}^{G+D-p+1}=0,\qd
s\om_0^{G+D}=0.\label{r3}\eeq
In particular \Gl{r3} contains \Gl{i1} i.e. the volume form
$\om_D^G$ contained in $f^{G+D}(\4C,\PH)$ solves \Gl{i1}.
 In order to realize that this part is given by \Gl{r0} one uses
\beq \4C^a=\Th^m\e ma,\qd \4C^\m=C^\m+\Th^m\cA m\m,\qd
     \Th^m=dx^m+C^m\label{r10}\eeq
which holds according to (\ref{g11}) and (\ref{b5}).
Written in terms of the $C^m$ and $C^\m$, $f^{G+D}(\4C,\PH)$ therefore depends
on the ghost $C^m$ and the differential $dx^m$ only via their
sum $\Th^m$. Since the $\Th$'s anticommute $f^{G+D}(\4C,\PH)$ contains
no piece of higher degree than $D$ in the $\Th$'s.
This implies that the volume form contained in $f^{G+D}(\4C,\PH)$
is given by
\bea \left[f^{G+D}(\4C^a,\4C^\m,\PH)\right]_D
=\left[f^{G+D}(e^a,C^\m+\ca^\m,\PH)\right]_D\label{r00}\eea
and thus leads to \Gl{r0}. Notice that $\om_D^G$ does not depend on the
ghosts of diffeomorphisms $C^m$ at all. This
holds in particular for $G=1$ and thus may indicate
that diffeomorphisms are not anomalous for the
theories investigated here. I stress that this result holds due to
our assumptions that the vielbein is invertible and that the manifold is
contractible. In fact, in noncontractible manifolds which allow for closed
but not exact 1--forms $\7\om_1=dx^m\om_m(x)$ there are generally
solutions of \Gl{i1} with $G=1$ which depend on the $C^m$ (and on the
$\7\om_1$) as will be shown
in section \ref{rem}. Notice that \Gl{r00} shows that $\om_D^G$ is
more conveniently written using the ghost basis $\{C^m,C^\m\}$ than using the
basis $\{\7C^N\}$. In contrast, the 0--form $\om_0^{G+D}$ is most
conveniently written in terms of the ghosts $\7C^N$ according to
lemma 1 of the next section since it is BRS-invariant:
\beq \om_0^{G+D}=f^{G+D}(\7C,\PH) .\label{rr1}\eeq
For all other forms occurring in the descent
equations there is no preferred ghost basis in which they take a
simple form since they depend both on
the differentials (via the gauge field 1--forms $\ca^N$)
and on the ghosts $C^m$.
\bsk

\noindent {\it Remarks:}

(i) Notice that \Gl{b5} and \Gl{g11} imply
\bea f^{G+D}(\7C,\PH)=
 \om(C^m,C^\m,\cA mN,\PH)\qd \then\qd
 f^{G+D}(\4C,\PH)=\om(\Th^m,C^\m,\cA mN,\PH).
\label{rr2}\eea
Thus we can describe our result also in the following way:
Modulo trivial contributions
each solution $\{\om_p^{G+D-p}:\ 0\leq p\leq D\}$ of the descent equations
can be obtained from
a BRS-invariant 0--form $\om_0^{G+D}=f^{G+D}(\7C,\PH)$ by expressing
the latter in terms of the ghosts $C^m$ and $C^\m$ and then replacing
each ghost $C^m$ by $\Th^m=dx^m+C^m$. The resulting function
$\om(\Th^m,C^\m,\cA mN,\PH)$ is the formal sum of all $\om_p^{G+D-p}$.
Such a result has been derived already in \cite{grav} for the special case of
Einstein--Yang--Mills theory. The derivation given in
\cite{grav} however uses the assumption that the background vielbein
is constant. The result of the present paper shows that this assumption is
superfluous.

(ii) According to (\ref{r0}) each solution of (\ref{i1}) with negative ghost
number is trivial since the r.h.s. of (\ref{r0}) does not depend on fields with
negative ghost number. The isomorphism of $H^{G,D}(s|d)$ and $H^{G+D}(\4s)$
implies then that (\ref{r1}) has nontrivial solutions only for $g\geq D$.
These statements are in fact not restricted to contractible manifolds.
Namely the vanishing of $H^{G,D}(s|d)$ for $G<0$ is a well-known consequence of
the simple BRS-transformation $s\zeta^N=b^N$ of the antighosts and can be
proved as the analogous statement in Yang--Mills theories (see e.g.
\cite{com}). The vanishing of $H^{g}(\4s)$ for $g<D$ is proved as follows:
A solution $f^g=\sum_p\om_p^{g-p}$ of \Gl{r1} with $g<D$ decomposes into the
descent equations \Gl{r3} with vanishing volume form $\om_D^{g-D}$
(recall that a function \Gl{r-1} does not contain parts of negative ghost
number). By means of arguments used at the end of section \ref{proof} one
concludes from $\om_D^{g-D}=0$ that in fact all forms $\om_p^{g-p}$ are
trivial and thus that $f^g$ is trivial.\msk

{\bf Corollary:} {\it $H^{g}(\4s)$
is zero for $g<D$ provided there are no constant ghosts.}

\mysection{Proof of the result}\label{proof}

In this section the result given in the previous section
is proved by means of the following lemmas:

{\bf Lemma 1:} \\
{\it a) Nontrivial contributions to local BRS--invariant forms can be chosen to
depend on the fields and their derivatives only via the undifferentiated ghosts
$\7C^N$ and via the tensor fields $\PH^r$:
\bea & & s \, \om(dx,x,\ph,\6\ph,\6\6\ph,\ldots)=0\nn\\
& \then& \om=f(dx,x,\7C,\PH)+
s\, \h(dx,x,\ph,\6\ph,\6\6\ph,\ldots).         \label{p0}\eea
b) A local function of the variables $dx^m,x^m,\7C^N,\PH^r$ is BRS--trivial
(in the space of local forms)
if and only if it is the BRS--variation of a local function of
the same variables:}
\bea & & \exists\, \h:\qd f(dx,x,\7C,\PH)=
        s\, \h(dx,x,\ph,\6\ph,\6\6\ph,\ldots)\nn\\
& \LRA & \exists\, g:\qd f(dx,x,\7C,\PH)=
s\, g(dx,x,\7C,\PH).\label{p0a}\eea
Lemma 1 is proved in appendix \ref{app1}.\msk

{\bf Lemma 2:}
{\it BRS-invariance of a function $\om(\7C,\PH)$ implies $\4s$-invariance
of the function $\om(\4C,\PH)$ which
arises from $\om(\7C,\PH)$ by replacing $\7C^N$ with $\4C^N=\7C^N+\ca^N$:}
\beq s\,\om(\7C,\PH)=0\qd \then\qd \4s\,\om(\4C,\PH)=0.\label{p1}\eeq
Lemma 2 holds since $\4s$ acts on the variables $\4C^N$ and $\PH^r$ exactly
as $s$ acts on the variables $\7C^N$ and $\PH^r$, see \Gl{ten9}, \Gl{t17},
\Gl{b4} and \Gl{b6}.\msk

{\bf Lemma 3:} {\it Algebraic Poincar\'e Lemma in contractible manifolds
and in absence of constant ghosts:
In the space of local forms
\[\om_p=dx^{m_1}\ldots dx^{m_p}\om_{m_1\ldots m_p}
(x,\ph,\6\ph,\6\6\ph,\ldots)\]
closed forms are also exact unless they are volume forms or constant 0--forms.
Volume forms are exact if and only if their Euler derivative
\[\0 {\7\6}{\7\6\ph^\a}=\sum_{n\geq 0}\qd \sum_{m_{r+1}\geq m_r}
(-)^{n}\6_{m_1}\ldots\6_{m_n}\0 {\6}{\6(\6_{m_1}\ldots\6_{m_n}\ph^\a)}\]
with respect to each elementary field $\ph^\a$ vanishes:}
\beq \begin{array}{rlcl}
p=0: & d\om_0=0 & \LRA & \om_0=const., \\
0<p<D: & d\om_p=0 & \LRA & \om_p=d\om_{p-1}, \\
p=D: & \om_D=d\om_{D-1} & \LRA &
\forall\ph^\a:\  {\7\6\om_D}/{\7\6\ph^\a}= 0.
\end{array}\label{p2}\eeq

For forms which do not depend explicitly on the coordinates but
only on the $\ph^\a$ and their partial derivatives this lemma has been
proved e.g. in \cite{bon,com,olv}.
In appendix \ref{app2}
this result is used to prove a generalized version of lemma 3.

We now turn to the proof of the result stated in the previous section.
In order to simplify the notation the ghost number of the forms is omitted in
the following.
Assume that $\om_D$ solves (\ref{i1}). Applying $s$ to (\ref{i1}) one
concludes by means of (\ref{i2}) that $s\om_{D-1}$ satisfies $d(s\om_{D-1})=0$.
Since $s\om_{D-1}$ is not a volume form this implies
according to (\ref{p2}) the existence of a local $(D-2)$--form
such that $s\om_{D-1}+d\om_{D-2}=0$. Repeating the argument one deduces
the existence of a set of local $p$--forms $\om_p$ which satisfy
\beq s\om_D+d\om_{D-1}=0,\qd s\om_{D-1}+d\om_{D-2}=0,\qd\ldots\qd ,\qd
s\om_{L+1}+d\om_{L}=0,\qd s\om_L=0.                           \label{gl7}\eeq
We shall see that in contractible manifolds
one actually has $L=0$, i.e. in this case the descent
equations (\ref{gl7}) terminate always with a nontrivial 0--form unless
$\om_D$ solves \Gl{i1} trivially (this does not hold for instance in
pure Yang--Mills theory unless one includes diffeomorphisms in $s$).
I first note that
the set of forms $\om_p$ is not unique since if the set $\{\om_p\}$
satifies eqs. (\ref{gl7}) then the set $\{\om'_p\}$ with $\om'_p=\om_p-s\h_p-
d\h_{p-1}$ also satisfies these equations
where $\{\h_p\}$ denotes an arbitrary set of local forms with $\gh(\h_p)=
\gh(\om_p)-1$. The sets $\{\om_p\}$ and $\{\om'_p\}$ are called
equivalent since their elements differ only by trivial contributions.
In particular, trivial contributions $s\h_L+d\h_{L-1}$ to $\om_L$
can be always absorbed by choosing an equivalent set. By means of (\ref{p0})
one therefore concludes that
without loss of generality $\om_L$ can be assumed to
depend on the fields and their derivatives only via the $\7C^N$ and $\PH^r$
\[ \om_L=\om_L(dx,x,\7C,\PH).\]
Evidently $\om_L$ can be written in the form
\beq \om_L=\sum_\tau \om^\tau(dx,x)f_\tau(\7C,\PH)\label{2}\eeq
where the $\om^\tau$ are linear independent $L$--forms which do not
depend on the fields and the $f_\tau$ are linear independent functions
of the variables $\7C^N$, $\PH^r$. Without loss of generality one
can further assume that no nontrivial linear combination of the $f_\tau$
combines to a BRS--trivial function:
\beq \sum_\tau \la^\tau f_\tau=s\cb \qd\LRA\qd \la^\tau=0\qd \forall\,\tau.
\label{2a}\eeq
Namely otherwise an appropriate trivial piece $s\h_L$ can be subtracted from
$\om_L$ such that (\ref{2a}) holds for $\om'_L=\om_L-s\h_L$
(cf. eq. (3.10) of \cite{grav} and the arguments used there).
$s\om_L$=0 requires
\beq  sf_\tau=0\qd\forall\, \tau\label{33}\eeq
due to the linear independence of the $\om^\tau$.
According to (\ref{p1})
each $f_\tau$ can be completed to a solution $\4f_\tau=f_\tau(\4C,\PH)$
of $\4s\4f_\tau=0$.
In particular this implies the existence of a 1--form $h_\tau$
whose BRS--transformation equals $-df_\tau$:
\beq  df_\tau+sh_\tau=0,\qd
h_\tau=\ca^N\0 \6{\6\7C^N}f_\tau(\7C,\PH). \label{4}\eeq
Inserting (\ref{2}) into the last but one eq. (\ref{gl7}) one obtains
(using (\ref{4}))
\beq \sum_\tau (d\om^\tau)\, f_\tau+s(\om_{L+1}-\sum_\tau \om^\tau h_\tau)=0.
\label{5}\eeq
If $d\om^\tau$ does not vanish for all $\tau$ then
(\ref{5}) requires that a nontrivial linear combination of the $f_\tau$ is
BRS-exact. This however contradicts (\ref{2a}) and therefore each of the
forms $\om^\tau$ has to be closed,
\beq d\om^\tau=0\qd \forall\, \tau.\label{6}\eeq
Furthermore the $\om^\tau$ can be assumed not to be exact,
\beq \om^\tau\neq d\h^\tau\qd\forall\, \tau\label{7}\eeq
since $ \om^\tau=d\h^\tau$ implies that the contribution $\om^\tau f_\tau$ to
$\om_L$ is trivial due to (\ref{4}) and therefore can be subtracted from
$\om_L$:
\[\om^\tau=d\h^\tau\qd \then\qd \om^\tau f_\tau=d(\h^\tau f_\tau)+s(\h^\tau
h_\tau)\qd (\mbox{no summation over $\tau$}).\]
Since we assumed the manifold to be contractible we conclude from (\ref{6})
and (\ref{7}) that $\om^\tau$ is a constant 0--form, i.e. in fact $\om_L$
is a 0--form depending only on the $\7C^N$ and $\PH^r$:
\beq \om_L=\om_0(\7C,\PH).\label{8}\eeq
Thus we have proved that to each nontrivial solution $\om_D$ of (\ref{i1})
there corresponds a solution $\om_0$ of
\beq s\, \om_0(\7C,\PH)=0,\qd \om_0\neq s\,\h_0(\7C,\PH).\label{rp1}\eeq
{}From (\ref{p1}) we also know that
conversely each nontrivial solution $\om_0$ of (\ref{rp1}) gives rise to
a solution $\4\om$ of \Gl{r1} which in fact is nontrivial since $\4\om=\4s
\4\h$ would imply the existence of a function $\h_0$ such that $s\h_0=\om_0$
in contradiction to \Gl{rp1}. Finally each nontrivial solution of \Gl{r1}
gives rise to a nontrivial solution $\om_D$ of \Gl{i1} according to the
following argument: Suppose that $\om_D$ is trivial,
i.e. that there are local forms $\h_D$ and $\h_{D-1}$ such that $\om_D=s
\h_D+d\h_{D-1}$. Inserting this into \Gl{i1} one obtains
$d(-s\h_{D-1}+\om_{D-1})=0$ which implies according to (\ref{p2}) the existence
of a local form $\h_{D-2}$ such that
$\om_{D-1}=s\h_{D-1}+d\h_{D-2}$. Thus the triviality of $\om_D$ implies
the triviality of $\om_{D-1}$.
Repeating the arguments one concludes that all other forms
$\om_p$ are trivial as well and thus that $\4\om=\4s\4\h$ which contradicts
the assumption that $\4\om$ is nontrivial. Thus the nontrivial solutions of
\Gl{i1} and \Gl{r1} correspond one-to-one which completes the proof.

\mysection{Discussion of some assumptions}\label{rem}

This section comments on three assumptions which have been made in order
to derive the results presented in sect. \ref{res}. These assumptions are
the contractibility of the manifold, the absence of constant
ghosts, and the algebraic independence of the ghosts and
their partial derivatives. For simplicity it is only discussed how the results
may change if only one of these assumptions is not fulfilled respectively.\msk

{\it a) Noncontractible manifold:}\msk

If the manifold is not contractible there may be non-exact solutions of
(\ref{6}) apart from constant 0--forms and thus the descent equations
\Gl{gl7} may
terminate with a form $\om_L$ whose degree $L$ differs from 0 and is
of the form
\beq \om_L= \sum_\tau \7\om^\tau_L(dx,x)f_\tau(\7C,\PH)\label{be5}\eeq
where $f_\tau(\7C,\PH)$ are BRS-invariant functions satisfying \Gl{2a}
and the $\7\om^\tau_L(dx,x)$ are closed but not exact $L$--forms
which can be chosen such that no nontrivial linear combination
of them combines to an exact form:
\beq d\7\om^\tau_L(dx,x)=0\qd \forall\tau,\qd
\sum_\tau\la^\tau\7\om^\tau_L(dx,x)=d\h(dx,x)\qd \LRA \qd
\la^\tau=0\qd \forall \tau.\label{be4}\eeq
One easily completes \Gl{be5} to an $\4s$-invariant function:
\beq \4\om=\sum_\tau\7\om^\tau_L(dx,x)f_\tau(\4C,\PH).\label{be5a}\eeq
$\4\om$ decomposes into parts of definite form degree and ghost number
which solve the descent equations
\Gl{gl7} and in particular contains a solution of \Gl{i1} which is given by
\beq \om^G_D=\sum_\tau \7\om^\tau_L(dx,x)\left[f_\tau(\4C,\PH)\right]_{D-L}
\label{be6}\eeq
where $\left[f_\tau(\4C,\PH)\right]_{D-L}$ denotes the $(D-L)$--form
contained in $f_\tau(\4C,\PH)$.
Notice that \Gl{be6}, unlike \Gl{r0},
generally depends on the ghosts of diffeomorphisms
$C^m$ even if one writes it in terms of the ghost basis $\{C^m,C^\m\}$
since $\4\om$ depends on $C^m$ and $dx^m$ not
only via their sum $\Th^m=C^m+dx^m$ unless $L=0$. I remark however that
the corollary given at the end of section \ref{res} implies that
without loss of generality
functions $f_\tau$ contributing to \Gl{be5} resp. \Gl{be5a} can be
assumed to have
degree $g\geq D$ in the $\7C^N$ resp. $\4C^N$ (provided there are no constant
ghosts). One easily verifies that this implies
\beq L=D+G-g\leq G\label{be6a}\eeq
i.e. only closed but not exact forms with degree $L\leq G$ can contribute
to a solution $\om_D^G$. In particular, apart from constant 0--forms, closed
but not exact forms can contribute to anomalies only if they are 1--forms.\msk

{\it b) Presence of constant ghosts}\msk

If the manifold is contractible the descent equations always terminate
at form--degree 0 (unless $\om_D$ is trivial) but in presence of constant
ghosts their general form is not given by \Gl{r3} but reads
\bea & & 0<p\leq D:\ s\om^{G+D-p}_p+d\om^{G+D-p+1}_{p-1}=0,\qd
s\om^{G+D}_0=\7\om^{G+D+1}_0(C_0),\nn\\
&\LRA& \4s\, f^{G+D}=\7\om^{G+D+1}_0(C_0),
\qd  f^{G+D}=\sum_{p=0}^D\om^{G+D-p}_p
\label{be7}\eea
where the possible occurrence of a nontrivial function $\7\om^{G+D+1}_0$
of the constant ghosts, denoted collectively by $C_0$,
complicates the situation.
A simple example shows that such a function really may occur.
We introduce the following set of fields: the
vielbein fields $\e ma$, a set of bosonic scalar fields
$\Ps^\m$, a corresponding set of constant fermionic ghosts $C_0^\m=C^\m$,
$\m=D,\ldots, 2D$ and the ghosts $C^m$ of diffeomorphisms. On these
variables we define the BRS--operator according to
\[ s\e ma=C^n\6_n\e ma+\e na\6_mC^n,\qd s\Ps^\m=C^m\6_m\Ps^\m+C^\m,\qd
sC^m=C^n\6_nC^m,\qd sC^\m=0.\]
I remark that this example can be formulated in the framework given in sect.
\ref{ass} according to
\beann & &[\cd_a,\cd_b]=\cf abc\cd_c,\qd [\cd_a,\De_\m]=[\De_\m,\De_\n]=0,
\qd \cf abc=\E am\E bn(\6_n\e mc-\6_m\e nc),\\
& &\cd_a\Ps^\m=\E am\6_m\Ps^\m,\qd \De_\n\Ps^\m=\de_\n^\m,\eeann
i.e. the set $\{\De_\m\}$ consists in this case of abelian generators
$\De_\m=\6/\6\Ps^\m$ which commute with the covariant derivatives and the
corresponding gauge fields $\cA m\m$ vanish. One can easily
verify that $\4s=s+d$ acts on the variables $\Ps^\m$ and $C^\m$ according to
\[ \4s\Ps^\m=F^\m+C^\m, \qd \4s C^\m=\4s F^\m=0, \qd
F^\m:=\Th^m\6_m\Ps^\m,\qd \Th^m=dx^m+C^m\]
(notice that one has $C^\m=\7C^\m=\4C^\m$ in this case).
Furthermore one can check that a solution of \Gl{be7} with $G=0$ is given by
\beann  \7\om_0^{D+1}&=&\ep_{\m_0\ldots \m_{D}}C^{\m_0}\ldots C^{\m_D},\\
f^D &=&\ep_{\m_0\ldots \m_{D}}\Ps^{\m_0}(C^{\m_1}\ldots C^{\m_D}-
F^{\m_1}C^{\m_2}\ldots C^{\m_D}                      \\
& &+F^{\m_1}F^{\m_2}C^{\m_3}\ldots C^{\m_D}-
\ldots+(-)^DF^{\m_1}\ldots F^{\m_D})                  \eeann
(all terms in $\4s f^D$ cancel
apart from $\ep_{\m_0\ldots \m_{D}}C^{\m_0}\ldots C^{\m_D}$
and $\ep_{\m_0\ldots \m_{D}}F^{\m_0}\ldots F^{\m_D}$; the latter vanishes since
it has degree $(D+1)$ in the $D$ anticommuting variables $\Th^m$).
\bsk

{\it c) Identities relating the ghosts:}\msk

The third assumption which is commented here is the assumption that the
ghosts are elementary fields (of course it does not matter whether one
regards the $\7C^N$ or the $C^m$, $C^\m$ as the basic ghost fields).
This assumption is needed for the proof of lemma 1 (see appendix \ref{app1})
and the following example shows that the results given in section \ref{res}
indeed generally do not hold if this assumption is not satisfied. We consider
$D$--dimensional gravity with Weyl transformations, i.e. the set
$\De_\m$ consists in this case of the generators $l_{ab}=-l_{ba}$
of Lorentz transformations and the generator $\de$ of Weyl transformations.
The algebra (\ref{a6}) is chosen torsionless, i.e. it is given by
\bea & &[\cd_a,\cd_b]=-\s0 12\R ab{cd}l_{cd}-F_{ab}\de,\qd
[\de,\cd_a]=-\cd_a,\qd
[l_{ab},\cd_c]=\h_{bc}\cd_a-\h_{ac}\cd_b,\nn\\
& & [l_{ab},l_{cd}]=2\h_{a[c}l_{d]b}-2\h_{b[c}l_{d]a},\qd
\qd [l_{ab},\de]=0\label{ex0}\eea
where $\h_{ab}=diag(1,-1,\ldots,-1)$ denotes the $D$-dimensional
Minkowski-metric, and
$\R ab{cd}$ and $F_{ab}$ denote the components of the
Riemann tensor and the field strength of Weyl--transformations.
They are obtained from \Gl{a13} which gives in this case
(with $\R mn{ab}=\e mc\e nd\R cd{ab}$, $F_{mn}=\e ma\e nb F_{ab}$)
\bea \R mn{ab}&=&\6_m\o n{ab}-\6_n\o m{ab}-\o {mc}a\o n{cb}+\o {nc}a\o m{cb},
\label{ex2}\\
F_{mn}&=& \6_mA_n-\6_nA_m,\label{ex3}\\
0&=&\6_m\e na-\6_n\e ma-\o {mb}a\e nb+\o {nb}a\e mb+\e ma A_n-\e na A_m
\label{ex4}\eea
where $\o m{ab}$ are the components of the spin-connection
(gauge field for Lorentz transformations) and $A_m$ denote the components
of the gauge field for Weyl transformations (indices $a$ are raised
and lowered by means of $\h_{ab}$).
(\ref{ex4}) expresses the vanishing of the torsion in $[\cd_a,\cd_b]$.
It can be solved for $\o m{ab}$ and thus determines it in terms of the
vielbein and $A_m$:
\bea \o m{ab}&=&E^{an}E^{br}(\om_{[mn]r}-\om_{[nr]m}+\om_{[rm]n}),\nn\\
\om_{[mn]r}&=&\s0 12e_{ra}(\6_{m}\e {n}a+\e {m}aA_{n}
-\6_{n}\e {m}a-\e {n}aA_{m})\label{ex5}.\eea
The covariant derivatives (\ref{a8})
take the form
\beq \cd_a=\E am(\6_m-\s0 12\o m{ab}l_{ab}-A_m\de)\label{ex1}\eeq
and the BRS--transformations \Gl{t23}--\Gl{t26} of $\e ma$,
$A_m$ and the ghosts read
\bea s\,\e ma&=&C^n\6_n\e ma+\e na\6_mC^n+C_b{}^a\e mb+C\e ma,\label{ex6}\\
s\, A_m&=&C^n\6_nA_m+A_n\6_mC^n+\6_mC,\label{ex7}\\
s\, C^m&=&C^n\6_nC^m,                     \label{ex8}\\
s\, C&=&C^n\6_nC,                        \label{ex9}\\
s\, C^{ab}&=&C^n\6_nC^{ab}+C_c{}^b C^{ac}\label{ex10}\eea
where $C^{ab}$ denote the Lorentz ghosts and $C$ is the Weyl ghost.
As long as $\et ma$, $A_m$, $C^m$, $C^{ab}$ and $C$ are elementary
fields the results given in section \ref{res} are valid. However there is
a particular choice of $C$ in terms of $C^m$ and $\e ma$ for which
(\ref{ex6})--(\ref{ex10}) are satisfied without dropping the assumption
that the remaining fields are elementary.
This choice is
\beq C=-\0 1{e\, D}\, \6_m(C^m\, e),\qd\ e=det(\e ma)\label{ex11}\eeq
One can check that the identification (\ref{ex11}) does not contradict
any of the assumptions given in section \ref{ass}
apart from the assumption that all ghosts are elementary
fields. In particular $A_m$ is still
an elementary field and not a function of $\et ma$ and its derivatives,
and the $\et ma$ are not subject to some identity
which restricts their independence (like, for instance,
$e=const.$).
As a consequence of (\ref{ex11}) there are local solutions
of (\ref{i1}) which are not obtainable by (\ref{r0}). They are
simply given by $d^Dx$ times an arbitrary function of $e$:
\beq \om_D^0=d^D\! x\, f(e).\label{ex12}\eeq
(\ref{ex12}) indeed solves (\ref{i1}) due to
\beq s\, e=0\label{ex13}\eeq
which can be explicitly verified using (\ref{ex6}) and (\ref{ex11}).
\Gl{ex12} is a nontrivial solution of \Gl{i1} since it is not a total
derivative (unless $f=const.$). The descent equations arising from
\Gl{ex12} just read $s\om_D^0=0$, i.e. they terminate at form-degree $D$,
not at degree 0. In particular there is no element of $H^D(\4s)$
which corresponds to \Gl{ex12}. Thus the results
given in section \ref{res} evidently do not hold in this case.
This shows that the assumption that the ghosts are elementary
fields is indeed decisive for the validity
of the results given in section \ref{res}. I remark that \Gl{ex6}--\Gl{ex11}
is the non-supersymmetric version of an analogous example
discussed in \cite{litim} for D=4, N=1 supergravity.

\mysection{Summary}\label{con}

It has been shown for the class of gauge theories described in section
\ref{ass} that each off-shell solution of the consisteny equation \Gl1 can
be obtained from a nontrivial solution of \Gl{r1}. This follows from the fact
that the descent equations \Gl{r3} which emerge from
the consisteny equation and which can be written
compactly in the form
\[ \4s\, f^{G+D}=0,\qd f^{G+D}=\sum_{p=0}^D\om_p^{G+D-p},\qd \4s=s+d\]
have a solution of the form
\[ f^{G+D}\cong\4C^{N_1}\ldots\4C^{N_{G+D}}f_{N_1\ldots N_{G+D}}(\PH)\]
where $\cong$ denotes equality up to trivial solutions of the form
$\4s g$. Thus the nontrivial part of $f^{G+D}$ can be written
as a function of the tensor fields $\PH$ given in \Gl{a5} and the
variables $\4C$ given in \Gl{b5} (resp. \Gl{r10}). The latter should be
considered as appropriate generalizations of the connection 1--forms
$\ca^N=dx^m\cA mN$.

This particular form of $f^{G+D}$ implies a remarkable statement about the
dependence of the solutions $\om_D^G$ of (\ref{i1}) on the various variables
(fields and their partial derivatives, explicit coordinates and
differentials). Namely according to \Gl{r0} $\om_D^G$ can be written,
up to trivial contributions, entirely in terms of the undifferentiated
ghosts $C^\m$, the 1--forms $e^a=dx^m\e ma$ and $\ca^\m=dx^m\cA m\m$ and
the tensor fields.
In particular the differentials and the gauge fields (including the
vielbein) contribute only in the combination of the 1--forms
$\ca^N\in\{e^a,\ca^\m\}$ apart from the dependence of the tensor fields
on the gauge fields. Since $\om_D^G$ is a volume form with ghost number $G$
it is of the form
\beq \om_D^G\cong\ca^{N_1} \ldots\ca^{N_D}C^{\m_1}\ldots C^{\m_G}\,
\Ga_{N_1\ldots N_D\m_1\ldots\m_G}(\PH).                  \label{c5}\eeq
If one converts the `world index' $m$ of $\cA m\m$ into a `flat index'
$a$ according to
\[ \cA a\m=\E am\cA m\m\qd \then\qd \ca^\m=e^a\cA a\m\]
then the differentials contribute in (\ref{c5}) only via the 1--forms $e^a$.
Using
\[ e^{a_1}\ldots e^{a_D}=d^D\!x\, e\, \ep^{a_1\ldots a_D},\qd
d^D\!x=dx^1\ldots dx^D,\qd e= det(\e ma),\qd \ep^{1\ldots D}=1\]
one therefore can write $\om_D^G$ as the volume element
$d^D\!x\, e$ times a function of the $C^\m$, $\cA a\m$ and $\PH^r$
whose degree in the $C^\m$ is fixed by its ghost number and which is a
polynomial of degree $D$ in the $\cA a\m$:
\beq \om_D^G\cong d^D\! x\, e\, \sum_{n=0}^DC^{\m_1} \ldots C^{\m_G}
\cA {a_1}{\n_1}\ldots\cA {a_n}{\n_n}\,
\om_{\n_1\ldots\n_n\, \m_1\ldots \m_G}^{a_1\ldots a_n}(\PH).\label{c6}\eeq
I stress that these results hold for arbitrary background metric $\bg ma(x)$
(provided it is invertible). In particular,
though we allowed $\om_D^G$ to depend arbitrarily (but locally) on $\et ma$ and
explicit coordinates, it has been shown that the solutions in fact depend on
these variables only via $\e ma=\bg ma(x)+\et ma$ and its derivatives up to
trivial contributions which of course can depend arbitrarily on $\et ma$ and
$x^m$. Furthermore the fact that $\om_D^G$ does not depend
on the ghosts of diffeomorphisms
may indicate that diffeomorphisms are not anomalous (provided
the vielbein is invertible).
I stress again that the results may change if the
 assumptions are weakend, see section \ref{rem}.
In particular
nontrivial solutions $\om_D^G$ of \Gl{i1} may depend
on the ghosts of diffeomorphisms if the manifold is not contractible
and possesses closed but not exact $p$--forms with $p\leq G$.
These forms then generally contribute also themselves
to $\om_D^G$ according to \Gl{be6}.
This does not affect the results for
classical actions since they have ghost number 0,
but there may be anomaly candidates which depend linearly on the
$C^m$ and on closed but not exact 1--forms
if the manifold allows for such 1--forms.
I remark that this holds also for Yang-Mills theories
in non-contractible manifolds as follows from the results of \cite{hen}
(recall that pure Yang--Mills theories are strictly speaking not covered by
the investigation of this paper unless one extends them to
Einstein--Yang--Mills theories).

One can check our result for known examples of solutions of (\ref{i1}). E.g.
in $D=4$ Einstein--Yang--Mills theory the solution of \Gl{i1}
corresponding to the nonabelian chiral anomaly can be written
in the following ways (see e.g. \cite{grav}):
\bea
\om^1_{4,chir}&=& Tr\left\{Cd(AdA+\s0 12 A^3)\right\}\label{ci1}\\
&=&Tr\left\{C(F^2-\s0 12A^2F-\s0 12AFA-\s0 12FA^2+\s0 12A^4)\right\}
\label{ci2}\\
&=&d^4\!x\, e\, \ep^{abcd}\, Tr\{C\, (F_{ab}F_{cd}
-\s0 12\ca_a\ca_bF_{cd}-\s0 12\ca_aF_{bc}\ca_d
-\s0 12F_{ab}\ca_c\ca_d+\s0 12\ca_a\ca_b\ca_c\ca_d)\}\label{ci3}\eea
with
\beann C=C^iT_i,\qd A=dx^m\cA mi T_i,\qd
\ca_a=\cA ai T_i,\qd F=dA+A^2=\s0 12 \, e^ae^b F_{ab}
\eeann
where $\f ijk$ are the structure constants of the
Liealgebra $\cg$ of the Yang--Mills gauge group,
$\{T_i\}$ is a set of constant matrices representing $\cg$ according to
$[T_i,T_j]=\f ijk T_k$ and $C^i$ and $\cA mi$ denote the Yang--Mills ghost
and gauge fields whose BRS--transformations \Gl{t26} resp. \Gl{t24} read
\[ sC^i=C^m\6_mC^i-\s0 12C^jC^k\f jki,\qd
s\cA mi=C^n\6_n\cA mi+\cA ni\6_mC^n+\6_mC^i-C^j\cA mk\f jki.\]
(\ref{ci2}) evidently is of the form (\ref{c5}) and
(\ref{ci3}) is of the form (\ref{c6}) since the entries of
$F_{ab}$ are tensor fields.
Furthermore one can verify that $\om_{4,chir}^1$ arises via (\ref{r0}) from
\[ f^5_{chir}(\4C,\PH)=Tr(\cC\cF^2-\s0 12\,\cC^3\cF+\s0 1{10}\,\cC^5),
\qd \cC=\4C^iT_i,\qd\cF=\s0 12\4C^a\4C^bF_{ab}\]
which solves (\ref{r1}) as one can verify easily by means of
$\4s\, \cC=-\cC^2+\cF$, $ \4s\, \cF=\cF \cC-\cC\cF$.
Namely one obtains $\4s f^5_{chir}(\4C,\PH)=Tr(\cF^3)$ which is a 6-form
in the anticommuting $\4C^a$ and therefore vanishes in four dimensions.
\vspace{1cm}

\appendix

\noindent{\LARGE {\bf Appendix}}

\mysection{Proof of Lemma 1}\label{app1}

In order to prove Lemma 1 of section \ref{proof}
we introduce variables which are better suited to this
problem than the original set of variables containing the elementary
fields and their partial derivatives. The new set of variables is
given by
\bea & &\{dx^m,x^m,\7C^N,\PH^r,u_l,v_l\},\qd v_l=s\, u_l,\nn\\
& &\{u_l\}=\{\et ma,\ \cA m\m,\ \zeta^N,\ \6_{(m_k}\ldots\6_{m_1}\cA {m_0)}N,
\ \6_{(m_k}\ldots\6_{m_1)}\zeta^N:
\ k\geq 1\}\label{p3b}\eea
where total symmetrization is defined according to
\[ T_{(m_1\ldots m_n)}=\0 1{n!}\sum_{\pi}
T_{m_{\pi(1)}\ldots m_{\pi(n)}}\]
($\pi$ runs over all permutations in the symmetric group $S_n$).
We claim that each local function of the variables
$dx^m,x^m,\ph^\a,\6_m\ph^\a,\6_m\6_n\ph^\a,\ldots$ can be also expressed
as a function of the variables (\ref{p3b}), i.e.
\beq \om(dx,x,\ph,\6\ph,\6\6\ph,\ldots)=
\Om(dx,x,\7C,\PH,u,v).\label{p3}\eeq
In order to prove (\ref{p3}) it is sufficient to show that each of the
variables $\6_{m_k}\ldots\6_{m_1}\ph^\a$
can be expressed as a local function of the variables (\ref{p3b}).
This can be shown by induction on the order $k$ of partial
derivatives of the $\ph^\a$.
For $k=0$ the statement is obviously true.
Let us denote an arbitrary $k$th order partial derivative
of one of the $\ph$'s by $\6^{(k)}\ph$ and
assume that the statement has been proved
up to order $k$, i.e. $\6^{(k)}\ph=\Om^{(k)}(x,\7C,\PH,u,v)$.
In order to prove it for order $k+1$ we consider
\bea \6_m\6^{(k)}\ph=\6_m\Om^{(k)}(x,C,\PH,u,v)
=\left\{\0 \6{\6x^m}+(\6_m\7C^N)\0 \6{\6\7C^N}+(\6_m\PH^r)\0 \6{\6\PH^r}
\right.\nn\\
\left.+(\6_mu_l)\0 \6{\6u_l}+(\6_mv_l)
\0 \6{\6v_l}\right\}\Om^{(k)}(x,C,\PH,u,v).
\label{p3a}\eea
Now one considers the various terms appearing on the r.h.s. of (\ref{p3a}).
According to (\ref{t18}) $\6_m\7C^N$ can be written as
$s\cA mN-\7C^M\cA mP\cf PMN$
which for all values of $N$ is a function of the
$x^m,\7C^N,\PH^r,u_l,v_l$ since
by assumption $\cf PMN=\cf PMN(\PH)$ and since $s\e ma=s\et ma$.
According to (\ref{ten9c}) $\6_m\PH^r$ can be written as $\cA mN\De_N\PH^r$
which is evidently a function of the variables $x^m,\PH^r,u_l$ since
by assumption $\De_N\PH^r$ is a function $R_N{}^r(\PH)$. Therefore
the first three contributions to the r.h.s. of (\ref{p3a}) can indeed
be written in terms of the variables (\ref{p3b}) but we still have
to show that $\6_mu_l$ and $\6_mv_l$ can also written in terms of
these variables. This is trivial if $u_l$ or $v_l$ denote
derivatives of $\zeta^N$ or $b^N$ and holds for the remaining $u$'s and
$v$'s if it holds for all derivatives of the $\ca$'s and $\7C$'s. To prove
it for all $\6^{(k+1)}\ca$ we use
\beq \6_{m_{k+1}}\ldots\6_{m_1}\cA {m_0}N
= \6_{(m_{k+1}}\ldots\6_{m_1}\cA {m_0)}N+\s0 {k+1}{k+2}
\6_{(m_1}\ldots\6_{m_k}{X_{m_{k+1})m_0}}^N\label{p3c}\eeq
where
\[{X_{mn}}^N=\6_{m}\cA {n}N-\6_{n}\cA {m}N.\]
According to (\ref{t19}) the
second term on the r.h.s. of (\ref{p3c})
is proportional to
\beann \6_{(m_1}\ldots\6_{m_k}(\cA {m_{k+1})}M\cA {m_0}P\cf PMN)\eeann
and can be expressed as a polynomial in the $\6^{(n)}\ca$, $n\leq k$
with coefficients which are functions of the $\PH^r$ (recall that
$\cf MNP=\cf MNP(\PH)$ and $\6_m\PH^r=\cA mN R_N{}^r(\PH)$).
Thus each partial derivative of $\cA mN$ of order $k+1$ can be expressed
in terms of the variables (\ref{p3b}) provided this holds also for
all lower orders and since it holds for $k=0$ it is true for all $k$.
In order to prove that $\6^{(k+1)}\7C$ can be expressed
in terms of the variables (\ref{p3b}) we use
\beq \6_{(m_1}\ldots\6_{m_{k+1})}\7C^N=s\,
\6_{(m_1}\ldots\6_{m_k}\cA {m_{k+1})}N-
\6_{(m_1}\ldots\6_{m_k}(\7C^N\cA {m_{k+1})}P\cf PMN).\label{p3d}\eeq
Since the second term on the r.h.s. depends only on those
$\6^{(n)}\7C$ with $n\leq k$ and on derivatives of $\ca$'s and $\PH$'s
the induction works also for the derivatives of the $\7C^N$.
This completes the proof of (\ref{p3}).

One now can apply standard techniques (for instance the Basic Lemma
\cite{let}): One introduces the `contracting' operator $r$ defined by
\beann r=\sum_lu_l\0 \6{\6v_l}                                   \eeann
whose anticommutator with $s$ is the counting operator $\cN$ for the
variables $u_l$ and $v_l$,
\beann \{r,s\}=\sum_l(v_l\0 \6{\6v_l}+u_l\0 \6{\6u_l})=:\cN,\eeann
decomposes a solution $\Om$ of $s\Om=0$ according to
$\Om=\sum_n \Om_n(dx,x,\7C,\PH,u,v)$ into
eigenfunctions $\Om_n$ of $\cN$ with eigenvalue $n$ and concludes that
each part $\Om_n$, $n\neq 0$ is BRS-exact. The locality of $\Om$
implies that $\Om_0$ does not depend on the $u_l$ and $v_l$ and this
proves lemma 1.

I stress that this proof takes advantage of the fact that there are no
algebraic identities between the $u$'s and $v$'s since otherwise
neither $\cN$ nor $r$ are well-defined. The decisive assumption
which guarantees the absence of such identities is the assumption that
the ghosts are independent elementary fields.
 Namely on the one hand this guarantees the absence of algebraic
 identities between the $v$'s and on the other hand it requires
also the absence of algebraic identities
containing the variables $u$'s since the BRS-variation of
each identity between the $u$'s would
imply an identity involving the $v$'s.

\mysection{Proof of an extension of the algebraic Poincar\'e lemma}\label{app2}

We prove a lemma which reduces to lemma 3 of section \ref{proof}
if the manifold is contractible and if there are no constant ghosts.
The latter are denoted by $C_0$ as in section \ref{rem}.
The non-constant ghosts are contained in $\{\ph^\a\}$.\msk

{\bf Lemma 4:} {\it Closed but not exact contributions to local forms
\[\om_p=dx^{m_1}\ldots dx^{m_p}\om_{m_1\ldots m_p}
(x,C_0,\ph,\6\ph,\6\6\ph,\ldots)\]
do not depend on the $\ph^\a$ and their partial derivatives unless $\om_p$
is a volume form with nonvanishing Euler derivative $\7\6/\7\6\ph^\a$
with respect to at least one $\ph^\a$:
\beq \begin{array}{rlcl}
p=0: & d\om_0=0 & \LRA & \om_0=\7\om_0(C_0), \\
0<p<D: & d\om_p=0 & \LRA & \om_p=d\om_{p-1}+\7\om_p(dx,x,C_0), \\
p=D:& \multicolumn{3}{l}{\om_D=d\om_{D-1}(dx,x,C_0,[\ph])+
d^D\, x\cl(x,C_0,[\ph])+\7\om_D(dx,x,C_0),}\\
& \multicolumn{3}{l}{\exists\,\a:\qd {\7\6\cl}/{\7\6\ph^\a}\neq 0.}
\end{array}\label{be2}\eeq
The closed but not exact parts $\7\om_p(dx,x,C_0)$ are linear combinations
of closed forms $\7\om^\tau_p(dx,x)$
with linear independent coefficients $f_\tau(C_0)$,
\beq \7\om_p(dx,x,C_0)=\sum_\tau f_\tau(C_0)\7\om^\tau_p(dx,x) ,
\qd d\7\om^\tau_p(dx,x)=0\label{be3}\eeq
which can be chosen such that}
\beq \sum_\tau\la^\tau\7\om^\tau_p(dx,x)=d\h(dx,x)\qd \LRA \qd
\la^\tau=0\qd \forall \tau.\label{be44}\eeq

Lemma 4 is an extension of an analogous result which holds
for forms depending locally on the
elementary fields $\ph^\a$ and their partial
derivatives but not explicitly on the coordinates. Using
the notation $[\ph]$ for
the variables (\ref{a1}) this result reads

{\it Algebraic Poincar\'e lemma for local forms $\om_p(dx,[\ph])$}
\cite{bon,olv,com}:
\beq\ba{rlcl}
p=0:& d\om_0([\ph])=0&\LRA&\om_0=const.,\\
0<p< D:& d\om_p(dx,[\ph])=0&\LRA&\om_p=d\om_{p-1}(dx,[\ph])+\7\om_p(dx),\\
p=D:& \om_D(dx,[\ph])=d\om_{D-1}(dx,[\ph])& \LRA&
{\7\6\om_D}/{\7\6\ph^\a}= 0\qd \forall\ph^\a.\ea      \label{ap1}\eeq
Of course the lemma holds analogously in presence of constant ghosts which
then just have to be added
to the arguments of all forms and functions appearing in (\ref{ap1}) in order
to get the correct version of the lemma for local forms
$\om_p(dx,C_0,[\ph])$. Using (\ref{ap1}) we now prove (\ref{be2})
by an inspection of the dependence of a closed form
$\om_p(dx,x,C_0,[\ph])$ on the
partial derivatives of the $\ph^\a$.
To this end we introduce the counting operator $\Npar$ of
the total number of partial derivatives acting on the $\ph^\a$. It
acts as follows:
\beann \Npar\, \6^{(r)}\ph^\a=r\, \6^{(r)}\ph^\a,
\qd \Npar\, (\6^{(r)}\ph^\a\6^{(s)}\ph^\be)
=(r+s)\, \6^{(r)}\ph^\a\6^{(s)}\ph^\be,\qd \ldots\ \eeann
where $\6^{(r)}\ph^\a$ denotes an arbitrary partial derivative of
$\ph^\a$ of $r$th
order, i.e. $\6^{(r)}\ph^\a\in\{\6_{m_1}\ldots\6_{m_r}\ph^\a\}$.
Each local form $\om_p(dx,x,C_0,[\ph])$
can be decomposed uniquely into
$\Npar$--eigenfunctions
$\om_p^{(n)}$ where $n$ denotes the $\Npar$--eigenvalue. The
locality of $\om_p$ guarantees that only nonnegative integers $n$ appear
in this decomposition and that there is
a largest eigenvalue which is denoted by $\on$:
\beq\om_p(dx,x,C_0,[\ph])=\sum_{n=0}^\on\om_p^{(n)}(dx,x,C_0,[\ph])
,\qd \Npar\, \om_p^{(n)}=n\, \om_p^{(n)}.\label{ap2}\eeq
$d$ is decomposed into a part $d^x$ which acts nontrivially
only on the variables $x^m$ and a part $d^\ph$ acting nontrivially
only on the variables $[\ph]$:
\beq\ba{lll} d^x\, dx^m=d^x\, C_0=0,& d^x\, x^m=dx^m,&
d^x\, \6^{(r)}\ph^\a=0,\\
 d^\ph\, dx^m=d^\ph\, C_0=0,& d^\ph\, x^m=0,&
d^\ph\, \6^{(r)}\ph^\a=dx^{m}\6_{m}\6^{(r)}\ph^\a.\ea
\label{ap3}\eeq
Evidently $d^x$ and $d^\ph$ satisfy $[\Npar,d^x]=0$, $[\Npar,d^\ph]=d^\ph$.
Therefore $d\om_p=0$ decomposes into the set of equations
\beq d^x\om_p^{(0)}=0,\qd 0\leq n< \on:\qd
d^\ph\om_p^{(n)}+d^x\om_p^{(n+1)}=0,\qd d^\ph\om_p^{(\on)}=0.\label{ap6}\eeq
Keeping in mind that $d^\ph$ treats the variables $x$ and $C_0$ as constants
one concludes
by means of (\ref{ap1}) from the last of the equations (\ref{ap6}) that
$\om_{p}^{(\on)}$ (i) is $d^\ph$--exact  or (ii)
is a volume form with nonvanishing
Euler derivative with respect to at least one $\ph^\a$ or (iii)
 does not depend on the $[\ph]$ at all. (i)
can hold only if $\on\neq 0$ and $p \neq 0$, (iii) only if
$\on=0$. Thus one has one of the following mutually excluding cases:
\beq \begin{array}{rll}\mbox{(i)}&  \on\neq 0,\ p \neq 0:
&\om_{p}^{(\on)}=d^\ph\om_{p-1}^{(\on-1)}(dx,x,C_0,[\ph]),\\
\mbox{(ii)} & p=D:&
\exists\,\a:\ \7\6\om_D^{(\on)}/\7\6\ph^\a\neq 0,\\
\mbox{(iii)}& \on= 0:&\om_{p}=\om_{p}(dx,x,C_0).\end{array}\label{ap7}\eeq
In the case (i) one considers the form $\om'_p$ defined by
\beq \mbox{(i)}:\qd \om'_p(dx,x,C_0,[\ph]):=\om_p(dx,x,C_0,[\ph])-d
\om_{p-1}^{(\on-1)}(dx,x,C_0,[\ph])\label{ap8}\eeq
where $\om_{p-1}^{(\on-1)}$ is the form appearing in (\ref{ap7}(i)).
(\ref{ap8}) states that $\om_p$ is exact up to
a closed form $\om'_p$ whose decomposition into $\Npar$--eigenfunctions
${\om'}_{p}^{\, (n)}$ contains only parts with eigenvalues
$n<\on$ as can be seen by inserting (\ref{ap2}) into (\ref{ap8}) taking into
account (\ref{ap7}(i)). In the case (ii) one analogously considers
the form
\beq\mbox{(ii)}:\qd \om'_D:=\om_D-\om_D^{(\on)}\label{ap9}\eeq
which also contains only
$\Npar$--eigenfunctions with eigenvalues smaller than $\on$.
By induction on $n$ starting from the highest eigenvalue $\on$ one
therefore can prove
\bea p=0: & &\om_0=\om_0(x,C_0),\nn\\
0<p<D: & & \om_p=d\om_{p-1}(dx,x,C_0,[\ph])+\7\om_p(dx,x,C_0),\nn\\
p=D:& & \om_D=d\om_{D-1}(dx,x,C_0,[\ph])+
d^D\, x\cl(x,C_0,[\ph])+\7\om_D(dx,x,C_0),\nn\\
& &\exists\,\a:\qd\7\6\cl/\7\6\ph^\a\neq 0\label{ap10}\eea
where $\om_{p-1}$, $\om_{D-1}$ and $\cl$ are local since they are finite
linear combinations of forms whose $\Npar$--eigenvalues do not exceed $\on$
as follows from the proof. Finally one easily verifies that $d\om_0=0$
requires $\om_0=\om_0(C_0)$ and that the forms
$\7\om_p(dx,x,C_0)$ appearing in (\ref{ap10}) for $p>0$ can be chosen
to satisfy (\ref{be44}) by absorbing exact contributions
into $\om_{p-1}$. This completes the proof of (\ref{be2})
but I add a remark about the global existence of $\om_{p-1}$.
It is assumed that $\om_p(dx,x,C_0,[\ph])$ exists
globally (on the whole manifold) with regard to its dependence on the $x^m$.
Together with $d\om_p=0$ this requires:\\
a) each $\om_{p}^{(n)}$ exists globally (in the same sense) since
the $\om_{p}^{(n)}$ are independent due to the assumption of the independence
of the variables (\ref{a1}) (cases where this assumption is not
justified are clearly beyond the scope of this paper),\\
b) due to a) each $d^\ph\om_{p}^{(n)}$ also exists
globally since $d^\ph$ does not change the $x$--dependence,\\
c) (\ref{ap6}) and b) shows that $d^x\om_{p}^{(n)}$ exists globally.\\
One conludes:\\
d) (\ref{ap7}(i)) shows that $\om_{p-1}^{(\on-1)}$ and
$d^x\om_{p-1}^{(\on-1)}$ exist globally since this
holds for $\om_{p}^{(\on)}$ and $d^x\om_{p}^{(\on)}$ according to a) and
c).\\
e) Due to a) and d), $\om'_p$ given in (\ref{ap8}) also exists
globally. The same holds of course for $\om'_D$ in (\ref{ap9}). \\
Since $\om'_p$ and $\om'_D$ contain only parts with
$\Npar$--eigenvalues smaller than $\on$ one proves by iteration of the
arguments that all forms occurring in (\ref{ap10}) exist globally.

The version (\ref{p2}) of (\ref{be2}) holding in contractible manifolds
follows from the fact that in this case each closed form $\7\om_p(C_0,dx,x)$
is exact apart from constant 0--forms $\7\om_0(C_0)$.

\end{document}